# A framework for assessing value and heterogeneity, illustrated using an early model of population screening with a multi-cancer early detection test

A framework for assessing value and heterogeneity in economic evaluation

Kunst N[1] (https://orcid.org/0000-0002-2409-4246), Dias S[2] https://orcid.org/0000-0002-2172-0221), Payne K[3] (https://orcid.org/0000-0002-3938-4350), Palmer S[1] (https://orcid.org/0000-0002-7268-2560), Soares MO[1]* (https://orcid.org/0000-0003-1579-8513)

1 Centre for Health Economics, University of York, York, UK

2 Centre for Reviews and Dissemination, University of York, York, UK

3 Manchester Centre for Health Economics, School of Health Sciences, The University of Manchester, Manchester, UK

*Corresponding author, marta.soares@york.ac.uk

**Acknowledgements:** We thank GRAIL for providing the Tafazzoli model.

**Funding disclaimer**: This project is funded by NHS England. The views expressed are those of the authors and not necessarily those of NHS England. Any errors are the responsibility of the authors. Katherine Payne is supported by the National Institute for Health and Care Research (NIHR) Manchester Biomedical Research Centre (BRC) (NIHR203308)

**Competing interests**: The authors have no competing interests to declare that are relevant to the content of this article.

**Authors contributions:** Kunst conducted the early model; Kunst, Palmer and Soares conceptualised the framework, Soares wrote the first draft of the manuscript; all authors contributed critically to the final manuscript and approved submission.

**Abstract:**

Introduction: We present a framework to assess the economic value of healthcare interventions by disaggregating value and examining heterogeneity. We applied it to an early health-economic model of population screening in England with a multi-cancer early detection (MCED) test, the Galleri® test. Value for such technologies often includes benefits, such as those associated with earlier detection, alongside potential harms from, for example, false positives or overdiagnosis. Value can also vary between individuals; for example, for Galleri it can vary between those identified with different cancer types and at different stages. Understanding value components and heterogeneity is crucial for assessing overall value and prioritising subsequent research.

Methods: We adapted an existing decision-analytic model to simulate annual Galleri screening for an asymptomatic 50-year-old cohort, evaluating health outcomes in Quality Adjusted Life Years (QALYs) from an English NHS perspective (£; year of 2024). We disaggregate headroom value (assuming no cost to the test) into key components: cancer identification (pre- and post-diagnosis), false positives, overdiagnosis, and misclassification. Value from cancer identification post diagnosis was further disaggregated by cancer type to explore heterogeneity.

Results: Our analysis predicts an overall estimated headroom value of 0.135 QALYs per individual. Early cancer identification was a major contributor, driven primarily by health-related effects rather than cost savings. Cancers of the colon and rectum, lung and ovary were the most significant contributors, accounting for 50% of the overall value. These results were generally robust to a range of scenario analyses.

Conclusion: The value disaggregation framework can guide decision making by supporting judgements over the plausibility of overall value estimates, by helping understand heterogeneity, and by helping prioritise further model and evidence development activities.

**Key points**:

- Standard economic evaluations often hide how health technologies generate value and how value is accrued across different diseases or patient groups.
- We developed a framework that breaks down overall net benefit into specific gains, harms, and by subgroups/diseases, illustrating its use on multi-cancer early detection (MCED) screening.
- Applied to the MCED test, the framework revealed that, in the economic model, test value is driven by early diagnosis in a small number of cancers, showing how this tool helps decision-makers identify key value drivers and prioritise future research.

## 1. Introduction

Evidence on the economic value of healthcare interventions supports the efficient allocation of resources. To support decision making, economic value is typically summarised using metrics like the incremental cost-effectiveness ratio (ICER) (compared to a cost-effectiveness threshold) or the incremental net benefit. Uncertainty is typically characterised, aiming to describe incomplete or limited evidence on parameter values or structural and temporal uncertainties. While economic value is often presented using a single figure, there are often different mechanisms by which a particular use of a technology or intervention affects value (in relation to comparators). These different components may contribute positively (net health benefits) or negatively (net health harms) to overall value. For drugs, components of value may be the direct health benefits of treatment (positive), adverse events and/or added treatment costs (negative). For diagnostic (in symptomatic populations) or screening (in asymptomatic populations) tests, value may arise from the information tests provide which may (indirectly) improve subsequent care, and from adverse events, overdiagnosis[1] or false positives cases[2]. Complex interventions, which often include multiple elements or target multiple behaviours [1], may be expected to have multiple health effects. Complexity in value accrual may also occur where there is heterogeneity, i.e. where value accrual differs significantly across individuals. A specific case for heterogeneity arises when the technology presents multi-disease impacts.

The balance of value components and the way these distribute across individuals determines the net value of the technology. Disaggregating value and examining the relative contribution of components/groups of individuals can promote a better understanding of how value is accrued, help assess the credibility of estimates, support the prioritisation of further research and, where relevant heterogeneity factors are knowable in advance of the decision, support decisions about restricting adoption to subgroups.

In this paper, we develop and apply a framework for examining value and heterogeneity. We demonstrate this framework using an early health-economic model of the Galleri® (GRAIL) multi-cancer early detection (MCED) test blood test for population screening of an asymptomatic population. The use of Galleri in this context could potentially lead to detection of cancers at earlier stages when treatment is likely to be more effective and perhaps less costly. The Galleri test identifies 21 different sites of origin (or cancer types), and its impacts are expected to differ by cancer type and stage. Beyond early detection, screening with Galleri is expected to be associated with other components of value, such as overdiagnosis. Our early modelling tool is used to disaggregate value, and we develop graphical displays to show the contribution of different components and the impact of heterogeneity, including net health benefits accrued across cancer-types.

[1] Overdiagnosis is here defined as the diagnosis, from screening, of a cancer that would not have been diagnosed under current care.
[2] False positives arise from imperfect test specificity, i.e. when there is some likelihood of the test being positive in individuals without (detectable) cancer.

## 2. Methods

We adapted an existing cost-effectiveness model [2] to examine the use of the Galleri test for population screening in England. This cohort model simulates 50-year-old asymptomatic individuals either receiving yearly screening with Galleri until the age of 79-years alongside current care, or current care alone. Its predictive approach, common to other GRAIL models. [3-5], calculates (predicts) the likelihood of cancer cases detected under current care being detectable earlier (whilst preclinical), using the interception approach developed by Hubbel et al [4], from evidence on cancer incidence (by type, stage, age, and sex), dwell time (by cancer type and stage)[3] and test sensitivity (by cancer type and stage)[4] to predict stage shift and associated timings. The model applies a common structure and assumptions, but distinct parameter values, across 19 cancer types.

To calculate mortality implications of the predicted stage shifts, the model assumes that current care cancer mortality by stage of diagnosis applies to screen-detected cancers, with no mortality during lead time. The use of a predictive approach for both stage-shifts and mortality arises from the absence of empirical data on either of these endpoints. Other model assumptions include: a four-stage preclinical cancer progression model, and constant and independent pre-clinical progression rates between stages. The model does not evaluate uncertainty over stage-shift or mortality predictions. Within-tumour heterogeneity is limited to age/sex distribution of clinical cancer diagnosis.

The adapted model considered Quality Adjusted Life Years (QALY) as the main outcome measure and adopted an English NHS perspective and 2024 costs (£). This ‘early’ model, considering estimates and assumptions that are preliminary, is not considered sufficiently robust to support decision making about the use of the test itself but provides an important basis to guide future modelling efforts and evidence development in support of future policy decisions.

### *2.1 Model adaptation*

The existing model, incorporating 19 categories of solid cancers, was adapted to the English context using cancer- and stage-specific incidence and mortality rates [5] (extracted on November 22, 2022, from the National Cancer Registration and Analysis Service and the Office for National Statistics).

Galleri’s cancer- and stage-specific sensitivity estimates were updated using a separate analysis described in full elsewhere [6]. This analysis, using evidence from Klein et al [7], explored heterogeneity by looking at whether the data supports the sharing of evidence across cancer types and stages. Results showed large variation in test sensitivity, particularly for early-stage cancers, indicating that there are some cancer types that the Galleri test is more likely to find early than others. Further examination also supported by evidence of ctDNA shedding in the literature, identified a group of “low sensitivity” cancers (bladder, kidney and thyroid), but the heterogeneity across other cancer types could not be explained further. To input in the early model, we used estimates from a synthesis model that constrained sensitivity estimates in each tumour type to increase with stage, and used a common hierarchical effect to describe stage-specific estimates across cancer types, except for the "low sensitivity" cancers. The hierarchical effect imposes sharing of information but does so flexibly to reflect the level of heterogeneity across cancer types. Because heterogeneity is high, the sharing imposed by the hierarchical model in this application is only weak. Estimates are presented in Supplementary Material (Section S1).

[3] Dwell time reflects time to stage progression given the cancer does not get clinically detected at that stage or the individual does not die from other causes at that stage.
[4] Test sensitivity reflects the likelihood of the Galleri test being positive for cancer, where cancer of a particular type and stage is present.

The discount rate was updated to 3.5%. [8] In this early analysis, we assumed 100% adherence to screening with Galleri to facilitate interpretation. The original model added costs and QALY impacts from current screening programmes (bowel; breast; cervical; lung) to time spent without cancer diagnosis, which leads to cost savings from early cancer detection with Galleri. However, uncertainties regarding Galleri's impact on eligibility for other programmes and concerns that the anticipated cost savings may not be fully realised, led us to exclude costs and QALY impacts of existing cancer screening programs from the early model.

The costs of misclassification of cancer type by Galleri (incorrect cancer signal of origin) and false positive (FP) work-up costs were derived from Hackshaw et al.[9] and treatment-related costs from Wills et al.[10]. All costs were adjusted to 2024 estimates using the Consumer Price Index (CPI). Estimates are presented in Supplementary Material (Section S2).

*2.2 Base case and scenarios*

In this early analysis, the reliance on evidence and assumptions that are preliminary means that the base case does not constitute a preferred set of assumptions. The results of the base case should therefore be interpreted alongside the extensive series of scenario analyses. To examine alternative assumptions over parameters that were updated from the original US model and over parameters with less robust evidence and modelling assumptions, we explored the following eleven scenarios: cancer treatment costs (scenario 1), level of misidentification of cancer type, with implications for costs (scenario 2), test sensitivity (scenarios 3-5), overdiagnosis rate (scenario 6). Scenario 7 excluded benefits for cancers with existing screening programmes (assumes 0% MCED test sensitivity). Table 1 and Supplementary Material (Section S3) present further detail on these.

<< Table 1 here>>

*2.3 Cost-effectiveness evidence: headroom value*

The early analysis determined 'headroom' value for the Galleri test, defined as the value realised if the test were free, i.e. we do not include the test price in our model. This approach can be applied throughout the different stages of the development cycle of a technology where the price is not yet established, allowing for subsequent analyses to determine the highest price at which a technology is of economic value. We here do not examine the potential price-ceiling as our main interest is to examine value components and value drivers to inform priorities for future model and evidence development.

The metric of (headroom) value used here is the net health benefit, NHB, expressed in QALYs. The net health benefit is a metric that rearranges the elements of the conventional cost-effectiveness equation in the following way:

$$NHB = Health - Cost \;/\; k \qquad (1)$$

Here, *k*, is the cost-effectiveness threshold (assumed to be £20,000 per QALY as per NICE guidance [8]. Incremental NHBs, or iNHB, are the difference between intervention and comparator NHBs. iNHB can also be expressed (Equation 2) as the incremental health (*iHealth*) subtracted by the opportunity costs, i.e. the health consequences of making other health care resources (un)available as a result of the additional costs, *iCosts*, imposed (or saved) with the strategy of interest. Opportunity costs are subtracted from direct health impacts to yield iNHB:

$$iNHB = iHealth - \underbrace{iCost \;/\; k}_{\textit{opportunity costs}} \qquad (2)$$

*2.4 Disaggregating value*

Our analysis disaggregates headroom value into the following components: cancer identification, false positives, overdiagnosis and misidentification of cancer type. Using the model, the iNHB is broken down into contributions from each component *c* (out of the total number of components, *C*), as follows:

$$iNHB = \sum_{c=1}^{C} iNHB_c. \qquad (3)$$

where $iNHB_c$ reflects the contribution of a particular component, *c*, to overall value, including both the probability of a component applying (i.e. not all individuals incur all value components), and its impact on the individual's net benefit. This assumes additive value components, meaning that each component's impact is distinct and the sum of impacts across components reflects the total impact on the individual. In the early model, additivity holds because components other than cancer identification (for example, false positives or overdiagnosis) only affects health-related quality of life and costs, not disease progression or mortality.

For cancer identification, we further disaggregate value by cancer type to explore heterogeneity. This can be generalised to any component, c, being disaggregated by any groups of individuals, *g* (out of a total number of groups, *G*):

$$iNHB_c = \sum_{g=1}^{G} iNHB_{c,g} \qquad (4)$$

Value from cancer identification is assumed additive across cancer types because the model does not consider multiple cancer diagnoses in the same individual or interactions between individuals. The way the model was parameterised meant that only time post-diagnosis was disaggregated by cancer type, but not time pre-diagnosis. For this reason, in our results, pre-diagnosis (reflecting differences in the time to diagnosis) is identified as a separate component of value. Formulas 3 and 4 can be generalised to:

$$iNHB = \sum_{c=1}^{C} \sum_{g=1}^{G} iNHB_{c,g}. \qquad (5)$$

We used stacked bar plots to display overall value and the contribution of each cancer type. Summaries of input parameters (such as test sensitivity) and of outputs (such as predicted stage-shift) were tabled to examine value drivers within- and between-cancer types.

## 3. Results

Table 2 presents the early modelling tool's predictions for the base case. Annual screening with Galleri in England is predicted to have an overall headroom value (excluding the cost of screening with the test) of 0.135 QALYs per individual, representing the maximum value of the Galleri screening program to the healthcare system. Value stems mostly from health impacts (0.136 QALY) and opportunity cost (Equation 2) impacts are minimal (0.001 QALY). More detailed results are presented in Supplementary Material (Section S4).

<<Table 2 here>>

Table 2 further disaggregates total headroom value by component (the sum across components retrieves overall value, demonstrating the additivity assumption). Negative contributions, though negligible to the total within the structure and assumptions of the early modelling tool, include false positives, overdiagnosis and misclassification. Time spent prior to cancer diagnosis (shorter under active screening with Galleri) also presents a negative impact, of -0.138 QALY, but this is offset by the (positive) contribution to overall value of post-diagnosis, of 0.278 QALY. By disaggregating post-diagnosis value by cancer type, we identify that:

- three cancer types – colon and rectum, lung and bronchus and ovarian cancers– account for 50% of the overall headroom iNHB,
- five cancer types – the above plus head and neck and lymphoma – account for 70% of the headroom iNHB, and
- ten cancer types – all of the above plus pancreas, prostate, breast HR positive, liver and intrahepatic, bile and duct and oesophagus cancers – account for 88% of the headroom iNHB.

This highlights significant heterogeneity in cancer type contribution to Galleri's overall value. To examine how heterogeneity arises, for each of the top 5 cancers, Figure 1 plots the absolute NHB of detection and treatment of cancer for a given stage at diagnosis (scatter plot, y-axis to the left) and the difference in numbers detected at different stages with Galleri, as predicted by the model (bar plot, y-axis to the right). The two highest value cancers – colon and rectum and lung and bronchus cancers – are predicted to have significant reductions in late-stage diagnosis (stages 3 and/or 4) with Galleri. Despite lung and bronchus cancers being predicted to have larger reduction in stage 4 cancer, colon and rectum cancers show a higher difference in NHB between early and late-stage diagnosis. Ovarian, head and neck, and lymphoma present lower changes in late-stage detection. From these three cancer types, ovarian shows the largest NHB impact from stage shifting, explaining its higher contribution (of the three) to the value of Galleri test. Further details are in Supplemental Material.

<< Figure 1 here>>

As emphasised previously, the base case analysis does not constitute a preferred set of assumptions but should be interpreted alongside the extensive series of scenario analyses examining alternative data sources and assumptions. Figure 2 presents an overview of results across all the scenario analyses, focusing on the contribution to value across cancer types. Specifically, in this figure: the top of each bar indicates the headroom value of Galleri; the overall height of each bar indicates the contribution of early detection and treatment of cancer (post-diagnosis) to overall headroom value for each scenario; the bar is further broken down to reflect the proportional contribution of the top 5 cancer types individually (starts at the bottom with the $1^{st}$ cancer type) and of the remainder which are pooled into a single group (at the top of the bar). Because the cancer types were ordered, the cancer-type breakpoints should be interpreted in terms of their proportional contribution to the total, instead of reading their y-values.

<< Figure 2 here>>

The results indicate that Galleri's overall headroom value (value of y-axis at the top of the bar) remains generally robust for the majority of the seven scenarios. The height of the bar, reflecting the value of cancer detection and treatment (post-diagnosis), varies across test sensitivity scenarios ([4, 5]) and when the potential benefits attributed to cancers with existing screening programmes are excluded ([7]). Despite these differences, the predicted relative contribution of individual cancer types remains similar across scenarios: the contributions of the top 5 cancers range from 58% to 71%, and of the top 10 cancers from 79% to 87% (Supplemental Material Table S8). This consistency suggests that the uncertainties in the parameterisation and scenarios considered do not significantly determine the rank order of cancer types in terms of value. In scenario 7, where the contributions of cancers with a current screening programme are removed (removing impacts of colon and rectum and lung and bronchus in the top 5), the contribution of the top 5 and 10 cancers to overall value are predicted at its lowest. Additional results of scenario analysis are in Supplementary Material (Section S4).

## 4. Discussion

In this work, we developed a framework that disaggregates value within an economic model. This provides a more granular understanding of value accrual and of the impact of heterogeneity. We applied this framework to an early model of the economic value of the Galleri MCED test when used for screening alongside existing screening programmes. The purpose of this ‘early analysis’ was not to inform an immediate policy decision, but to guide further model and evidence development. Commissioning and further research decisions are expected when the NHS-Galleri clinical trial[12], ongoing in England, reports its primary stage-shift endpoint. We expect the framework here developed to be applied to the model incorporating NHS-Galleri Trial data. A review of MCED models conducted as part of a separate stream of work identified and summarised a range of published models and critically reviews them to identify the need for further model and evidence developments.[13]

Our framework reveals the complexity of value accrual for MCED technologies. It helps to:

- *Identify the contribution of different elements to overall value:* The early model predicts that overdiagnosis, false positives and misidentification have a negligible impact on overall value. The plausibility of these estimates should be ascertained, and further work should consider how to best evidence and quantify these elements.
- *Identify the cancer types contributing the most to overall value.* The framework provides a clear basis for characterising heterogeneity. The top five cancers -- colon and rectum, lung, ovarian, head and neck, and lymphoma cancers -- contribute over 60% to overall value. This breakdown allows for explicit judgements on the accuracy and uncertainty over individual estimates. For example, as a targeted lung cancer screening programme is currently being implemented in the UK, the predicted value of Galleri from stage shifts in lung cancer (contribution to iNHB of 0.047 QALYs) may not be fully realised.
- *Examines the impact of scenarios on components of value:* Given the lack of empirical evidence for the Galleri test's clinical impact in an asymptomatic population, the limited differences in value estimates between scenarios suggest that the current model may not fully characterise uncertainty, i.e. that the main structural assumptions may be too strong [13]. Future modelling efforts must aim to appropriately reflect uncertainty (including structural uncertainty).

Our framework was applied to an existing (adapted) model and was therefore conditioned by the existing implementation of the model. For example, in our application, the time pre-diagnosis could not be disaggregated by cancer type. This highlights the importance of considering value disaggregation early in model development. Further extension of the framework is required that incorporates uncertainty using probabilistic sensitivity analysis. As with any quantitative form of analyses, our framework is only able to examine parameterised elements and scenarios. Policy decisions, however, will need to consider a broader set of evidence and judgements, even if qualitatively, for example, deliberation on the potential impact of Galleri on adherence to existing screening programmes.

Our work shows that where value accrual is complex, the framework of value disaggregation here developed can guide decision making, supporting judgements over the plausibility of overall value estimates and the need for further model developments and evidence. This is particularly important for technologies expected to be associated with multiple components of value (like diagnostic or screening tests, or complex interventions [1]) and those for which effects are expected to be heterogeneous, such as the example examined here of MCED tests impacting on multiple diseases.

Table 1: Specification of scenario analyses for the early modelling tool.

| Category | Base case [scenario 0] | Scenario analysis |
|---|---|---|
| Treatment costs | Based on estimates from Wills [10] | **[scenario 1]** Ratio comparing UK and US healthcare spending [11] applied to Tafazzoli's base-case estimates [2]. |
| Cancer Signal of Origin misidentification (implications for costs) | Level of misdiagnosis based on separate analysis | **[scenario 2]** Tafazzoli's base-case estimates [2]. |
| Detection from MCED test | MCED test sensitivity based on separate analyses in Appendix 2, 0% sensitivity assumed for category 'Other'. | **[scenario 3]** Tafazzoli's base-case estimates [2].<br>**[scenario 4]** 15% reduction of base-case estimates.<br>**[scenario 5]** 50% reduction of base-case estimates |
| Overdiagnosis | Overdiagnosis is 5% of mortality, as in Tafazzoli [2]. | **[scenario 6]** Overdiagnosis is 10% of mortality. |
| Detection with Galleri possible on cancers with current screening programmes? | Yes. | **[scenario 7]** Excluding benefits on cancers that currently have screening programmes (lung, colon and rectum, breast, and cervix cancers), i.e., 0% test sensitivity assumed for these cancers. |

Table 2: Headroom value of the Galleri test (price of the test not included): disaggregation of value and of heterogeneity using the early modelling tool, base case.

| | | **Headroom value** (excludes cost of screening) threshold of £20,000/QALY | | |
|---|---|---|---|---|
| | | **iNHB** | **Incremental Health** | **Opportunity costs+** |
| | **Overall headroom value of Galleri** | 0.135 | 0.136 | 0.001 |
| | **Component of headroom value** | **iNHB** | **Incremental Health** | **Opportunity costs+** |
| | False positives | -0.005 | -0.002 | 0.003 |
| | Overdiagnosis | -0.002 | -0.001 | 0.001 |
| | Misidentification of cancer type | -0.000* | -0.000** | 0.000 |
| | Value arising from time spent prior to cancer diagnosis | -0.138 | -0.137 | 0.000 |
| | Early cancer detection and treatment (post-diagnosis) | 0.278 | 0.276 | -0.003 |
| | **Contribution of individual cancer types to value of early cancer detection and treatment** | **iNHB** (cumulative % contribution) | **Incremental Health** | **Opportunity costs++** |
| 1 | Colon and rectum | 0.059 (21%) | 0.057 | -0.002 |
| 2 | Lung and bronchus | 0.047 (38%) | 0.051 | 0.004 |
| 3 | Ovarian | 0.038 (52%) | 0.035 | -0.003 |
| 4 | Head and neck | 0.031 (63%) | 0.030 | -0.001 |
| 5 | Lymphoma | 0.019 (70%) | 0.019 | -0.001 |
| 6 | Pancreas | 0.012 (74%) | 0.013 | 0.001 |
| 7 | Prostate | 0.011 (78%) | 0.011 | 0.000 |
| 8 | Breast HR positive | 0.011 (82%) | 0.011 | 0.000 |
| 9 | Liver and intrahepatic bile duct | 0.009 (85%) | 0.010 | 0.001 |
| 10 | Oesophagus | 0.009 (88%) | 0.008 | 0.000 |
| 11 | Uterus | 0.007 (91%) | 0.007 | 0.000 |
| 12 | Stomach | 0.006 (93%) | 0.006 | 0.000 |
| 13 | Cervix | 0.005 (95%) | 0.005 | 0.000 |
| 14 | Bladder | 0.004 (96%) | 0.004 | 0.000 |
| 15 | Breast HR negative | 0.003 (98%) | 0.003 | 0.000 |
| 16 | Kidney and renal pelvis | 0.003 (99%) | 0.003 | 0.000 |
| 17 | Urothelial | 0.002 (99%) | 0.002 | 0.000 |
| 18 | Anus | 0.001 (100%) | 0.001 | 0.000 |
| 19 | Other+ | 0.000 (100%) | 0.000 | 0.000 |

+ Includes gallbladder, melanoma, sarcoma, thyroid cancers and other solid tumours.
++ Opportunity costs are calculated by multiplying the incremental costs by the assumed cost-effectiveness threshold (of £20,000 per QALY). These reflect the health consequences of making other health care resources (un)available as a result of the additional costs imposed (or saved) with the strategy of interest.
* $-3.589 \times 10^{-4}$
** $-1.397 \times 10^{-4}$

**Fig. 1**: Model inputs and estimates by cancer type for top 5 cancers, base case: absolute net health benefit (NHB) of cancer treatment by stage, and predicted change in stage-detection with Galleri

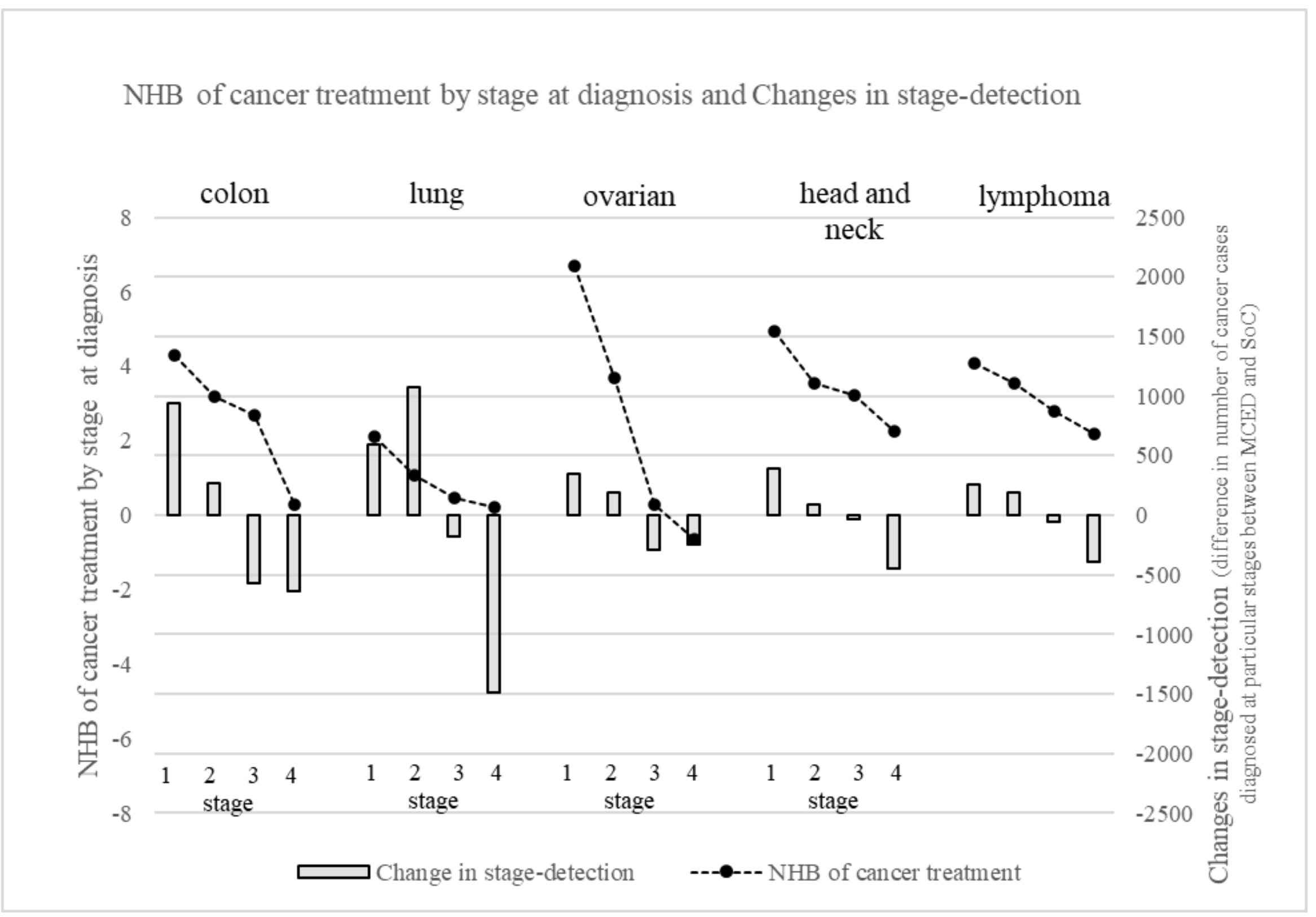

**Fig. 2**: Headroom value of the Galleri test (price of the test not included), highlighting the contribution of early detection and treatment of cancer (post-diagnosis) disaggregated for selected cancer types (the top 5 in the base-case: 1st – colon and rectum, 2nd – lung and bronchus, 3rd – ovarian, 4th – head and neck, 5th – lymphoma). Base case [scenario 0] and scenario analyses [scenarios 1-7].

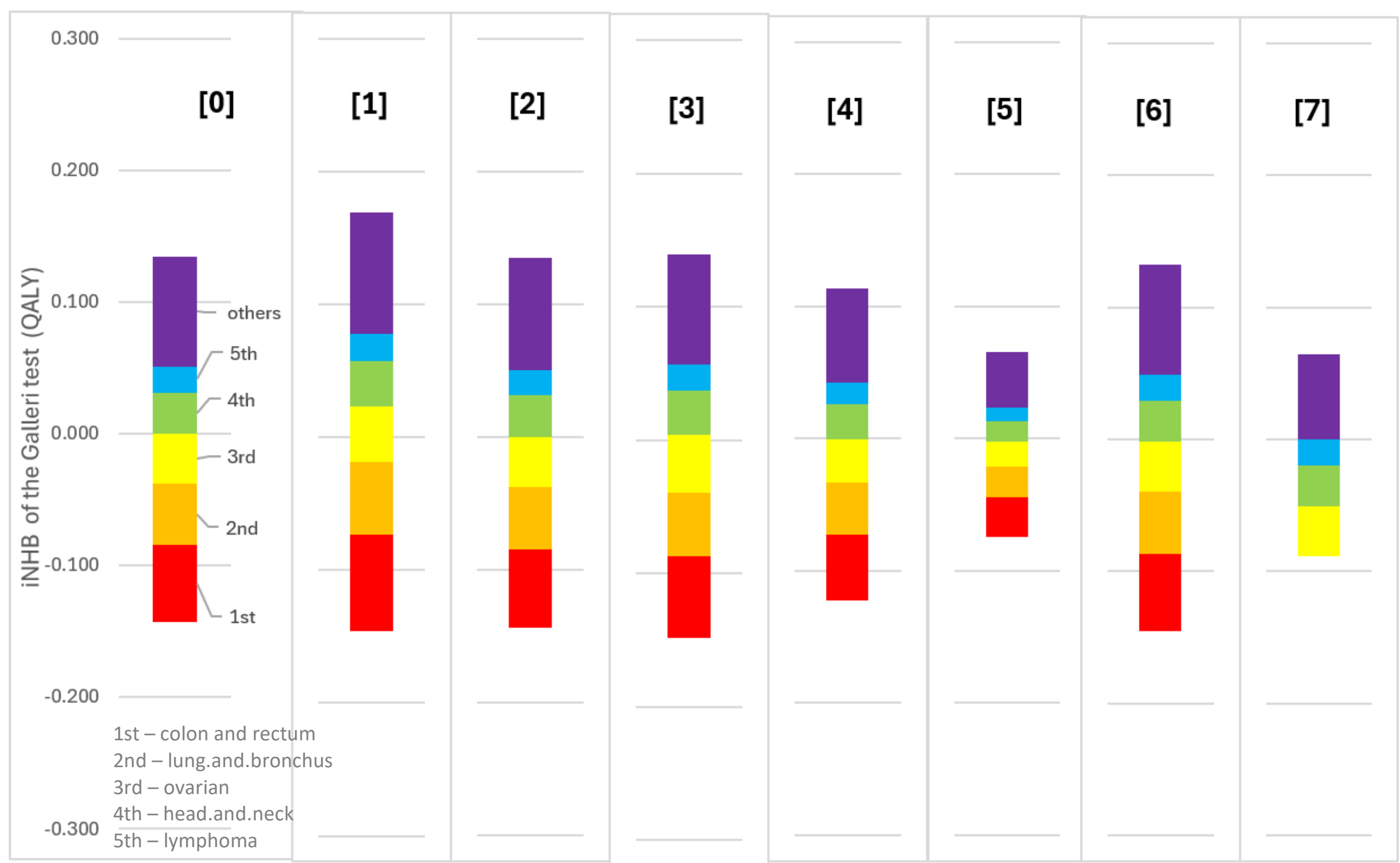

iNHB: incremental net health benefit; Scenarios (full specification in Table 1): [scenario 0] Base case, [scenario 1] Scenario on treatment costs, [scenario 2] Scenario on cancer signal of origin misidentification (implications for costs), [scenario 3] Tafazzoli's detection estimates, [scenario 4] 15% reduction of base-case sensitivity estimates, [scenario 5] 50% reduction of base-case sensitivity estimates, [scenario 6] Overdiagnosis is 10% of mortality, [scenario 7] Exclusion of benefits for cancers that currently have screening programme in England.